\documentclass{amsart}
\usepackage{amssymb, amsmath}

\newtheorem{thm}{Theorem}[section]
\newtheorem{lemma}[thm]{Lemma}

\newtheorem{cor}[thm]{Corollary}

\theoremstyle{definition}

\def\F{I\kern-.30em{F}}
\def\P{I\kern-.30em{P}}
\def\E{I\kern-.30em{E}}

\newcommand{\R}{\mathbb{R}}

\newcommand{\Z}{\mathbb{Z}}

\newcommand{\C}{\mathbb{C}}

\newcommand{\beq}{\begin{equation}}
\newcommand{\eeq}{\end{equation}}
\newcommand{\bea}{\begin{eqnarray}}
\newcommand{\eea}{\end{eqnarray}}

\newtheorem{lem:peter}{Lemma}

\def \Real {{\mathbb R}}

\def \Integers {{\mathbb Z}}
\title[Schr{\"o}dinger Operators with Random vector Potentials]{Localization
for Schr{\"o}dinger Operators \\
with Random Vector Potentials}

   \author { F.\ Ghribi, P.\ D.\ Hislop, and F.\ Klopp}
\thanks{PDH partially
supported by NSF grant DMS 0503784. FK partially
supported by Institut Universitaire de France.}
\begin{document}

\begin{abstract}

We prove Anderson localization at the internal band-edges for
periodic magnetic Schr{\"o}dinger operators perturbed by random vector
potentials of Anderson-type. This is achieved by combining new
results on the Lifshitz tails behavior of the integrated density of
states for random magnetic Schr{\"o}dinger operators, thereby
providing the initial length-scale estimate, and a Wegner estimate,
for such models.
\end{abstract}
\maketitle



\section{Introduction}
\label{introduction1}

Random magnetic Schr{\"o}dinger operators have attracted much recent
interest. These operators are technically challenging because the
randomness enters the coefficients of first-order differential
operators unlike the zero-order case of Schr{\"o}dinger operators with
random electrostatic potentials. This means that the variation of
the eigenvalues of the finite-volume operators is not monotone with
respect to the random variables. In this paper, we treat some new
models and prove localization near the internal band-edges due to a
random perturbation of the vector potential only. We consider random
vector potentials $A_\omega (x)$ of Anderson-type having the form
\beq \label{potential1} A_\omega (x) = \sum_{j \in \Integers^d}
\omega_j u(x-j), \eeq where $\{ \omega_j ~| ~j \in \Integers^d \}$ is
a family of independent, identically distributed $(iid)$ random
variables, and the single-site vector potential $u$ is a real,
vector-valued function of compact support. Such a random vector
potential generates a random magnetic field $B_\omega = d A_\omega$.
Precise hypotheses are formulated below. We consider families of
magnetic Schr{\"o}dinger operators on $L^2 ( \Real^d)$ obtained
by random perturbations of periodic magnetic Schr{\"o}dinger operators
by a random vector potential (\ref{potential1}). Such random
Schr{\"o}dinger operators have the form \beq \label{pert1}
H_\omega(\lambda) \equiv ( i \nabla + A_0 + \lambda A_\omega )^2 +
V_0 , \eeq where $V_0$ is a real-valued, bounded, periodic
electrostatic potential and $A_0$ is a real, bounded,
periodic vector potential. The unperturbed operator is $H_0 \equiv (
i \nabla + A_0 )^2 + V_0$.
For the disorder $\lambda > 0$ sufficiently small, the deterministic
spectrum $\Sigma (\lambda )$ has a band structure. We prove that
under certain hypotheses, and for the disorder $\lambda$
sufficiently small, there is a neighborhood of any internal band
edge in the deterministic spectrum $\Sigma (\lambda )$ that is
purely pure point spectrum with probability one, with exponentially
decaying eigenfunctions.


There are few results on localization for random magnetic
Schr{\"o}dinger operators. Nakamura \cite{[N1],[N2]} considered the
Lifshitz tails behavior of the integrated density of states (IDS)
$N(E)$ at the bottom of the spectrum for a general family of random
vector potentials associated with random magnetic fields on the
lattice $\Z^2$ and on the continuum $\R^d$, respectively. For models
$H_\omega = (i \nabla + A_\omega)^2$ on $L^2 ( \R^d)$, for $d \geq
2$, Nakamura \cite{[N2]} proves an upper bound on $N(E)$ as $E
\rightarrow 0^+$. Since zero is the bottom of the deterministic
spectrum provided zero is in the support of the distribution of the
random variables, this upper bound on the decay of the IDS provides
an initial length scale estimate for the multiscale analysis for an
interval of energies near zero. In order to prove localization in
this interval, it is also necessary to have a Wegner estimate near
zero energy. However, since zero is not a fluctuation boundary, no
such Wegner estimate is currently known.

The situation for lattice models is better. Nakamura \cite{[N1]}
proved a Lifshitz tail behavior at zero energy for a discrete model
in two-dimensions. The magnetic Hamiltonian is defined using a
vector potential $A((x,y))$, defined on the edges $(x,y)$, if $|x-y|
= 1$, and has the form
\beq\label{discrete1} (H \psi) (x) = \sum_{y: |x-y| =1} ( \psi(x) -
e^{i A ((x,y)) } \psi (y) ) ,
\eeq
for $u \in \ell^2 ( \Z^2)$. The vector potential satisfies $A((y,x))
= - A((x,y))$. The spectrum of this operator is $[0,8]$. The
magnetic field $B$ is defined for each unit square $F$ by $B(F) =
\sum_{e \in \partial F} A(e)$. Nakamura assumes that the collection
$B(F)$ is a family of independent, identically distributed random
variables. He proves that the IDS exhibits Lifshitz tails as $E
\rightarrow 0$. This calculation was extended by Klopp, Nakamura,
Nakano, and Nomura \cite{[KNNN]} who considered a specific model for
which $E_0 \equiv \inf \Sigma$ can be computed. They prove that the
IDS has Lifshitz tails as $E \rightarrow E_0$. These authors also
prove a Wegner estimate for this model. Using these two results,
Klopp, Nakamura, Nakano, and
Nomura \cite{[KNNN]} proved exponential localization near the bottom
of the spectrum for this family of lattice models in two-dimensions. In a
preprint,
Nakano \cite{[Nakano1]} proved localization for lattice models of the form
\beq\label{discrete2}
(H \psi) (x) = \sum_{y: |x-y| =1} e^{i A_\omega ((x,y)) } \psi (y)  ,
\eeq
where $\{ A_\omega \} \in [ - \pi, \pi)$ is a family of independent and
uniformly
distributed random variables. The deterministic spectrum
is $\Sigma = [ - 2d, 2d]$. Nakano proves that if the dimension $d \geq 11$,
then
there is a $\delta > 0$ so that the model exhibits Anderson localization on
$[-2d, -2d + \delta ] \cup [2d - \delta , 2d]$ with exponentially decaying
eigenfunctions.

Ueki \cite{[U0],[U1]} also consider random magnetic Schr{\"o}dinger
operators of the form (\ref{pert1}). In \cite{[U0]}, he studied
Lifshitz tails at zero energy for models on $L^2 ( \R^2)$ for which
the vector potential is constructed from a Gaussian random field. Ueki
studied localization for a similar random vector potential model on
$L^2 ( \R^d)$ in \cite{[U1]}, but there is an additional a random
electrostatic potential of Anderson-type. There, localization at
band-edges is due to the random electrostatic potential, rather than
the random vector potential. In this paper, we prove that a random
vector potential alone can create pure point spectrum almost surely in
neighborhoods of the band edges.

In order to state our main theorem, we list the necessary hypotheses
on our model.

\begin{enumerate}
\item[(H1)]
The bounded, real-valued function $V_0$ and the bounded, vector-valued
function $A_0$ are $\Z^d$-periodic. The self-adjoint operator $H_0 =
(i \nabla + A_0)^2 + V_0$ is $\Z^d$-periodic and essentially
self-adjoint on $C_0^\infty (\R^d)$. The semibounded operator $H_0$
has an open {\it internal} spectral gap. That is, there exist constants $-
\infty <
M_0 <  C_0 \leq E_- < E_+ < C_1 \leq \infty $ so that $\sigma
(H_0) \subset [ M_0 , \infty )$, and
$$
\sigma( H_0 ) \cap ( C_0 , C_1 )  = ( C_0 , E_- ] \cup [ E_+ , C_1
).
$$

\item[(H2)] Each component of the single-site vector potential $u_k$ is
continuously differentiable and compactly supported, i.\ e.\
  $u_k \in C_0^1 ( \R^d )$.  For each $k = 1, \ldots, d$, there
  exists a nonempty open set $B_k$ containing so that the
  single-site potential $u_k \neq 0$ on $B_k$.

\item[(H3)] The probability distribution of $\omega_0$ is absolutely
  continuous with respect to Lebesgue measure. The density $h_0$ has
  compact support contained, say, in $[-1 , 1 ]$ and the infimum of
  its support is negative and the supremum is positive. The density
  $h_0$ is assumed to be locally absolutely continuous.
\end{enumerate}

The periodicity of $H_0$ and the construction of the random vector
potential (\ref{potential1}) satisfying (H2)--(H3) imply that
$H_\omega (\lambda)$ is a $\Z^d$-ergodic of the family of operators.
As a consequence, there is closed subset $\Sigma (\lambda ) \subset
\Real$, the deterministic spectrum of $H_\omega ( \lambda)$, such that
$\sigma (H_\omega (\lambda) ) = \Sigma (\lambda)$ with probability
one. Furthermore, there are closed subsets $\Sigma_X (\lambda) \subset
\Sigma (\lambda)$, for $X= pp, ac, sc$ that are the pure point,
absolutely continuous, and singular continuous components of the
spectrum with probability one. Finally, it is known that the
deterministic spectrum $\Sigma (\lambda )$ has an open spectral gap
$G(\lambda) \equiv (E_- (\lambda) , E_+ (\lambda) ) \subset G = (E_- ,
E_+ )$ for $\lambda$ sufficiently small (cf.\ \cite{[BC]}). We note
that there are examples \cite{[HH],[N0]} of magnetic periodic
Schr{\"o}dinger operators ($V_0 = 0$) with open spectral gaps. We need
$V_0 \neq 0$ in order to construct examples for which condition (Gh)
(see section~\ref{internaltails1}) is satisfied.

\begin{thm}\label{maintheorem1}
  Suppose that $H_0$ is a periodic magnetic Schr{\"o}dinger operator
  satisfying the condition (H1), and that $H_\omega ( \lambda)$ is
  defined as in (\ref{pert1}), with Anderson-type random vector
  potential given in (\ref{potential1}) satisfying (H2)--(H3).
  Furthermore, we suppose that hypotheses (H4)--(H5) and condition
  (Gh) at both gap edges (see section \ref{internaltails1}) are
  satisfied.  There is a $\lambda_0 >0 $ such that for any $\lambda
  \in ( 0 , \lambda_0]$, there is an $\eta (\lambda) >0$, so that $E_+
  (\lambda) + \eta (\lambda) < E_+$ and $E_- < E_- (\lambda) - \eta
  (\lambda)$, and the deterministic spectrum in $I(\lambda) \equiv
  [E_- (\lambda) - \eta(\lambda), E_-(\lambda) ] \cup [ E_+(\lambda) ,
  E_+ (\lambda) + \eta (\lambda) ] $ is purely pure point with
  exponentially decaying eigenfunctions.  That is, we have that
  \beq\label{spectrum1} \Sigma (\lambda) \cap I(\lambda) = \Sigma_{pp}
  (\lambda) \cap I(\lambda), \eeq and there is no absolutely
  continuous or singular continuous spectrum in $I(\lambda)$:
  \beq\label{spectrum2} \Sigma_X (\lambda) \cap I(\lambda) =
  \emptyset, ~~X = \mbox{ac}, ~X = \mbox{sc}.  \eeq
%
%
%
%
%
%
%
%
\end{thm}

\noindent
It is clear that we can replace $\Z^d$ by a nondegenerate lattice
$\Gamma \subset \Z^d$, but we will only consider $\Z^d$ here.

To our knowledge, this is the first result on localization due to
only a random vector potential for continuum models. We will prove
Theorem \ref{maintheorem1} by combining the recent results of F.\
Ghribi \cite{[G1],[G2]} on Lifshitz tails for the IDS of
$H_\omega(\lambda)$ at internal band edges and the result of
\cite{[HK]} on the Wegner estimate.

There are now many results on localization for random
Schr\"odinger operators on $L^2 ( \R^d)$. Papers using multiscale analysis
include \cite{[BCH],[CH],[GK0],[K1],[KSS]},
and recently, the fractional moment method was extended to these models \cite{[AENSS]}.


\section{Internal Lifshitz Tails for the IDS}
\label{internaltails1}

We now discuss the main points of Ghribi \cite{[G1],[G2]}
on the Lifshitz tails behavior of the IDS $N(E)$ at the inner
band-edges of the deterministic spectrum $\Sigma(\lambda)$ of
$H_\omega ( \lambda)$. We recall from section \ref{introduction1}
that we have a $\Z^d$-periodic operator $H_0  = (i \nabla +
A_0)^2 + V_0$, on $L^2 ( \R^d)$, where $V_0$ is a real, bounded,
$\Z^d$-periodic, electrostatic potential, and $A_0$ is a real,
bounded, vector-valued $\Z^d$-periodic potential.
We denote by $C_0 \subset \R^d$ the unit cell and by $C_0^*$ the dual cell.
Since $H_0$ is periodic, it admits a Floquet decomposition. We denote
the Floquet eigenvalues by $E_n ( \theta )$,
for $ \theta \in C_0^*$,
and the corresponding normalized eigenfunctions are denoted by $\phi_{0, n}
( \theta, x)$.
We make some additional hypotheses on $H_0$.

\begin{enumerate}
\item[(H4)] The edge of the spectrum $E_+ $ is {\it simple} meaning
that it is attained by a single Floquet eigenvalue $E_{n_0} ( \theta)$, for
$\theta \in C_0^*$.
\item[(H5)] The IDS $N_0 (E)$ of $H_0$ at $E_+$ is {\it
nondegenerate} which means that: \beq\label{dos1} \lim_{\delta
\rightarrow 0^+} \frac{ \log ( N_0 ( E_+ + \delta) - N_0 ( E_+ ))}{
\delta} = \frac{d}{2} . \eeq
\end{enumerate}

Concerning (H4), we let $E_{n_0} (\theta)$ be the unique
Floquet eigenvalue that attains the band edge $E_+$.
Then, as proved in
\cite{[K2]}, there is a finite set of points $\theta_k \in C_0^*$ so
that $E_{n_0} (\theta_k) = E_+$, for $k = 1, \ldots, m$.
%

Let us suppose that $H_0$ and $H_\omega (\lambda)$ satisfy
hypotheses (H1)--(H5). There is an important hypothesis on $A_0$ and
$u$, necessary for the proof of Lifshitz tails,  that we call condition
(Gh).

\begin{enumerate}
\item[(Gh)] The matrix $M \equiv ( M_{k k'})_{1 \leq k, k' \leq m}$, with
matrix elements given by \beq\label{matrixele1} M_{k k'} = \int_{C_0}
~(( u \cdot i \nabla + i \nabla \cdot u + 2 u \cdot A_0 )
\phi_{0, n_0} )( \theta_k, x) \overline{\phi}_{0, n_0} (\theta_{k'},
x) ~dx , \eeq is positive or negative definite.
\end{enumerate}

\noindent
The following theorem follows from the main result of \cite{[G1]}.

\begin{thm}\cite{[G1]}
  \label{ids1} Assume that $H_0$ and $H_\omega (\lambda)$ satisfy
  hypotheses (H1)--(H5), and that condition $(Gh)$ is satisfied. Then,
  there exists $\lambda_0 > 0$ and $\nu > 0$ so that for all $\lambda
  \in [0, \lambda_0]$, one has $E_+ - \nu \lambda \leq E_+ (\lambda) \leq E_+$,
  and the IDS $N_\lambda (E)$ satisfies \beq \label{tails1} \lim_{E
    \rightarrow E_+ (\lambda)^+} \frac{ \log | \log (N_\lambda (E) -
    N_\lambda (E_+ (\lambda))) | }{ \log ( E - E_+ (\lambda)) } = -
  \frac{d}{2}.  \eeq A similar statement holds at the lower band edge
  $E_- (\lambda)$.
\end{thm}

We prove in section \ref{example1} that given a nontrivial $\Z^d$-periodic
potential $V_0$, we can construct a $\Z^d$-periodic vector
potential $A_0$, so that, with $\epsilon >0$ sufficiently small,
there are vector-valued functions $u$, and hence random
Anderson-type vector potentials (\ref{potential1}), so that
condition (Gh) is satisfied with $A_0$ replaced by $\epsilon A_0$.
We comment further on condition (Gh) in
section \ref{example1}.

Concerning hypothesis (H4), Klopp and Ralston \cite{[KR]}
proved that generically the band edge of a periodic Schr{\"o}dinger operator
is obtained by a single Floquet eigenvalue.
One expects that the same result should hold for a dense, open set of
pairs $(A_0, V_0)$ in $L^\infty ( C_0)$.
It is known that hypothesis (H5) is true in the one-dimensional case,
and not necessary for two-dimensional periodic
Schr{\"o}dinger operators \cite{[KW]}. That is, the band edge is always
nondegenerate.
Furthermore, hypothesis (H5) is known to be satisfied
for many families of periodic Schr{\"o}dinger operators, see the discussion in
\cite{[K2]}.


\section{Initial Estimate on the Resolvent of the Localized Hamiltonian}
\label{initial1}

We follow the method of using the Lifshitz tails behavior of the
IDS in order to obtain an initial decay estimate on the localized
resolvent of the local Hamiltonian obtained from the Hamiltonian by
restricting to finite-volumes
with periodic boundary conditions.
This method has been
used, for example, by Klopp \cite{[K1]}, in order to prove localization at
the bottom
of the deterministic spectrum, and in Klopp-Wolff \cite{[KW]}
in order to localization near internal
band edges.
The methods of \cite{[BCH]} or \cite{[KSS]} do not work for the models
of random vector potentials considered here because the variation of the
eigenvalues with respect to the random variables is not monotonic.
We localize the Hamiltonian $H_0$
to cube $\Lambda \subset \R^d$, with integer length sides, and obtain a
self-adjoint
operator $H_\Lambda^0$ by imposing periodic boundary conditions. It follows
from Floquet theory that the spectrum of $H_\Lambda^0$ still has a spectral
gap containing $G$.
We localize the random perturbation to such a region by writing
\beq\label{localpert1}
A_\Lambda (x) = \sum_{j \in \tilde{\Lambda}} \omega_j u(x-j).
\eeq
We assume that $A_\Lambda (x)$ is supported in $\Lambda$ without loss of
generality, which
amounts to the hypothesis that $\mbox{supp} ~u \subset C_0$, the unit cube.
We write $H_\Lambda (\lambda) = [(i \nabla + A_0 + \lambda A_\Lambda )^2 +
V_0] | \Lambda$,
with periodic boundary conditions.
The IDS for the local Hamiltonian $H_\Lambda ( \lambda)$ is defined as
\beq\label{loc-ids1}
N_\Lambda (E) = \frac{1}{| \Lambda |} \# \{ \mbox{eigenvalues of} ~H_\Lambda
( \lambda ) \leq E \} .
\eeq
For $\lambda$ sufficiently small,
this operator $H_\Lambda ( \lambda)$ also has an open spectral gap that
contains $(E_- (\lambda) , E_+ (\lambda))$.
We write $N_\Lambda (E)$ for the IDS for the operator $H_\Lambda (\lambda)$.

Theorem \ref{ids1} states that
the density of states in a small interval around the band-edge is
small. For example, it follows from Theorem \ref{ids1} that for any
$n \geq 1$, we have
\beq\label{dos2}
\lim_{E \rightarrow
E_+(\lambda)^+} (E - E_+(\lambda))^{-n}[ N(E)- N(E_+(\lambda))] = 0.
\eeq
The problem is to obtain information about the finite volume IDS
from this infinite volume result. For this, we use a result of Klopp and
Wolff
\cite{[KW]}.

\begin{thm}\label{kw1}
For any $E_0 \in \R$, if
\beq\label{kw2}
N(E_0 + \epsilon) - N(E_0 - \epsilon) = \mathcal{O} (\epsilon^\infty),
\eeq
as $\epsilon \rightarrow 0^+$, then for any $k \in \Z^+$, $k \geq 2$, any
$\nu > 0$,
and for any $L \in \Z^+$ sufficiently large,
\beq\label{kw3}
\E \left( N_{\Lambda_{L^k}}( E_0 + L^{-1}) - N_{\Lambda_{L^k}} (E_0 -
L^{-1}) \right)
\leq L^{-\nu} .
\eeq
\end{thm}

We apply this vanishing result as follows. We work with the upper band edge
$E_+ ( \lambda)$, a similar
argument holds for the lower band edge. Let $\chi_B ( \cdot )$ denote the
characteristic
function for the subset $B \subset \R$. We fix $\lambda > 0$, and consider
an
interval $I_\lambda (L) = [E_+ (\lambda), E_+ (\lambda) + L^{-1/2} ]$
near the upper band edge $E_+ (\lambda)$, for integer $L^{1/2}$ sufficiently
large, depending on $\lambda$,
so that $E_+ (\lambda) + L^{-1/2} < E_+$.
We apply Theorem \ref{kw1}, taking $k = 2$, and $\nu > 0$ arbitrary.
By the Chebychev inequality, we have
\bea\label{dos4}
\P \{ \sigma (H_{\Lambda_L}) \cap I_\lambda (L) \neq
\emptyset \}
&\leq & \P \{ Tr_{\Lambda_L} ( \chi_{I_\lambda (L) } (H_{\Lambda_L} ) ) \geq
1 \} \nonumber \\
&\leq & \E \{ N_{\Lambda_L} ( I_\lambda (L) ) \}  \nonumber \\
& \leq &
\E \left( N_{\Lambda_{L}}( E_+ (\lambda) + L^{-1/2}) - N_{\Lambda_{L}} (E_+
(\lambda) - L^{-1/2}) \right)
\nonumber \\
  & \leq  & L^{-\nu}.
\eea
%
%
%
So we are assured that the interval $[ E_+(\lambda) , E_+(\lambda) +
(4L)^{-1/2} ]$ is separated from the spectrum of
$H_{\Lambda_L}$ by an amount $(4 L)^{-1/2}$ with a probability
greater than $1 - L^{-\nu}$ according to the right side of (\ref{dos4}).
We use this as input into
the Combes-Thomas estimate \cite{[BCH],[CT],[GK]} on the exponential decay
of the localized
resolvent of $H_{\Lambda_L} (\lambda)$. For a cube $\Lambda \subset \R^d$,
we let $g_\Lambda$ be a smooth characteristic function of $\Lambda$ such
that
$g_\Lambda = 1$ except near a fixed neighborhood of the boundary $\partial
\Lambda$.
We will often write $g_L$ when $\Lambda$ is a cube of side length $L \in
\Z^+$.
We denote by $W(g_L)$ the commutator $W(g_L) = [ H_{\Lambda_L}, g_L]$
that is a first-order, relatively $H_0$-bounded operator localized near
$\partial \Lambda$.


\begin{thm}
\label{initialest1}
Fix $\lambda > 0$.
There is an $L_0 >>0$ and a $\gamma_0 > 0$, depending on $\lambda > 0$,
with $\gamma_0 L_0 \sim \mathcal{O} (L_0^{1/2}) >>1$, so that
for any energy $E \in [ E_+ (\lambda) , E_+ (\lambda) + (4 L_0)^{-1/2} ]$,
the local Hamiltonian $H_{\Lambda_{L_0}} (\lambda) $ satisfies the following
bound
\beq
\label{resolvent1}
\| W ( g_{L_0} ) R_{\Lambda_{L_0}} ( E ) \chi_{\Lambda_{L_0}} \| \leq e^{-
\gamma_0 L_0 },
\eeq
with a probability greater than $1 - {L_0}^{- \xi}$, for any $\xi > 2d$.
\end{thm}


\section{The Wegner Estimate for Random Magnetic Schr{\"o}dinger Operators}
\label{wegner1}

A Wegner estimate for random magnetic Schr{\"o}dinger operators of the
form (\ref{pert1}) with vector potentials of Anderson-type
(\ref{potential1}) was proven in \cite{[HK]} for energies in the spectral
gaps of $H_0$.
The proof of the
Wegner estimate differs from the usual proofs because the variation
in the eigenvalues with respect to the random variables is not
monotonic as it is in the random potential case. We recall the main
argument here. As in section \ref{initial1},
we localize our magnetic Schr{\"o}dinger operators (\ref{pert1}) to integer
side length cubes
$\Lambda_L$ with periodic boundary conditions. Expanding the form of the
operator $H_\Lambda (\lambda)$,
it will be convenient to write the local operator as
\beq\label{wegop1}
H_\Lambda ( \lambda) = H_0^\Lambda + \lambda H_1^\Lambda
+ \lambda^2 H_2^\Lambda,
\eeq
where $H_0^\Lambda = [(i \nabla + A_0)^2 + V_0]_{| \Lambda}$, with
periodic boundary conditions, is the fixed background operator with an
open spectral gap containing the open spectral gap $G$ of $H_0$.  As
the perturbation is $\lambda A_\Lambda$, we see that
\beq\label{pert-1}
H_1^\Lambda = [(i \nabla + A_0) \cdot A_\Lambda + A_\Lambda \cdot (i \nabla
+ A_0)]_{| \Lambda} ,
~~\mbox{and} ~~H_2^\Lambda = A_\Lambda \cdot A_\Lambda ,
\eeq
with periodic boundary conditions.

\begin{thm}\label{wegnerest1}
  Suppose that the deterministic background operator $H_0$ satisfies
  hypothesis (H1), and that the random operators $H_{j}^\Lambda, j =
  1,2$, defined in (\ref{pert-1}), are constructed with single-site
  vector potential $u$ and iid random variables satisfying (H2)--(H3).
  Suppose $G = (E_- , E_+ )$ is an open gap in the spectrum of $H_0$.
  Then, there exists a constant $ \lambda_0 >0$, and, for any $q > 1$,
  a finite constant $C_W > 0$, independent of $\lambda$, such that
  for all $| \lambda| < \lambda_0$, $E_0\in G$ and $\eta>0$ such that
  \begin{equation*}
    \frac{\lambda^2}{\text{dist} \; ( E_0 , \sigma ( H_0 )
      )}\leq\lambda_0,\quad [ E_0 - 2
      \eta , E_0 + 2 \eta ] \subset G.
    \end{equation*}
  we have
  \begin{equation}
    \label{wegner-1} \P
  \{ \; \text{dist} \; ( \sigma ( H_\Lambda ( \lambda )), E_0 ) \leq
  \eta \} \; \leq \; \frac{C_W}{\text{dist} \; ( E_0 , \sigma ( H_0 )
  )} \; \eta^{1/q} \; | \Lambda| .
  \end{equation}
\end{thm}

\vspace{.1in}
\noindent {\bf Remark.} As we will show in section
\ref{wegner1}, $ \text{dist}\; ( E_0 , \sigma ( H_0 )) \sim
\mathcal{O}( \lambda)$, so that the ratio in theorem
\ref{wegnerest1} is effectively $|\lambda| < \lambda_0$.

\vspace{.1in}

\begin{proof}

\noindent
1. We recall the basic ideas from \cite{[HK]}. Let $R_\Lambda (z) \equiv
(H_\Lambda (\lambda) - z)^{-1}$ be the
resolvent of the local operator $H_\Lambda (\lambda)$
on the Hilbert space $L^2 ( \Lambda)$. We begin with the observation
\beq\label{prob1}
\P \{ \mbox{dist} ( \sigma ( H_\Lambda (\lambda) ) , E_0 ) < \eta \}
   =  \P  \{ \; \| R_\Lambda ( E_0 ) \| > 1 / \eta \}
\eeq
Because we are working at the band-edges of an internal gap,
we use the Feshbach projection method to express the resolvent $R_\Lambda (
E_0 )$
in terms of various positive operators by reducing to
the spectral subspace of $H_0^\Lambda$
above $E_+$ and below $E_-$ (recall that the spectrum of $H_0^\Lambda$
always maintains the spectral gap).
Let $P_\pm$ be the spectral projectors for $H_0^\Lambda$ corresponding to
the
spectral subspaces $[ E_+ , \infty)$ and $( - \infty , E_-]$,
respectively. We consider the case of the upper band edge so that
$E_0 \in G$ and $E_0 > ( E_+ + E_- ) / 2$. The argument for the
lower band edge is similar. We will suppress $\Lambda$ for notational
simplicity,
and write $H_0^\pm  \equiv P_\pm H_0^\Lambda$, and
denote by $H_\pm ( \lambda ) \equiv H_0^\pm + \lambda P_\pm (\lambda
H_1^\Lambda + \lambda^2 H_2^\Lambda) P_\pm $.
We will need the various projections of operators $A$ between the
subspaces $P_\pm L^2 ( \R^d ) $, and we denote them by $A_\pm \equiv
P_\pm  A P_\pm $, and $A_{+-} \equiv P_+ A P_-$, with $A_{-+} =
A_{+-}^* = P_- A P_+ $. Let $z \in \C$, with $\mbox{Im} z \neq 0$.
We can write the resolvent $R_\Lambda ( z )$ in terms of
the resolvents of the projected operators
$H_\pm ( \lambda )$ as follows. In order to write a formula valid
for either $P_+$ or for $P_-$, we let $P = P_\pm $, $Q = 1- P_\pm $,
and write $R_P (z) = ( PH_0^\Lambda + P (\lambda H_{1 , \omega}^\Lambda +
\lambda^2 H_{2 , \omega}^\Lambda) P - zP )^{-1}$. We write the
effective perturbation as $\mathcal{V}(\lambda) \equiv \lambda H_1^\Lambda +
\lambda^2
H_2^\Lambda$. We then have
\beq\label{feshbach1}
R_\Lambda (z) = P R_P ( z ) P  +
\{ Q - P R_P (z) P \mathcal{V}(\lambda) Q \}  \mathcal{G} (z)
         \{ Q -  Q  \mathcal{V}(\lambda) P R_P (z)^* P \} ,
\eeq
where the operator $\mathcal{G}(z)$ is given by
\beq\label{feshbach2}
\mathcal{ G }(z) = \{  Q  H_0 + Q
  \mathcal{V} (\lambda) Q
- zQ - Q \mathcal{V} (\lambda)  P R_P ( z ) P \mathcal{V} (\lambda)
  Q \}^{-1} .
\eeq

\noindent
2. Our first goal is to reduce the estimate on the resolvent on the right in
(\ref{prob1}) to an estimate on the operator $\mathcal{G}(E_0)$.
Since we are working close to $E_+$, we take $P = P_-$ and $Q = P_+$.
We let $\delta_\pm (E_0) = \mbox{dist} ~( E_0 , E_\pm) = \mbox{dist} \; (
\sigma ( H_0^\pm ) , E_0 )$.
To this end, we note that
the resulting formula for the
first term on the right in
(\ref{feshbach1}), $P R_P (E_0) P \equiv R_- (E_0)$,
is
\bea\label{feshbach3}
R_- (E_0) & = & R_0^- (E_0)^{1/2}
     \{ 1 + R_0^- (E_0)^{1/2} P_- ( \lambda H_1^\Lambda
  \nonumber \\
  &  & +  \lambda^2  H_2^\Lambda  ) P_- R_0^- (E_0)^{1/2} \}^{-1}
    R_0^- (E_0)^{1/2},
\eea
provided the inverse exists. The first factor on the right in
(\ref{feshbach3}) exists provided $| \lambda | < \lambda_0^{(1)}$,
where $\lambda_0^{(1)}$ is fixed by the requirement that
\beq\label{lambda1}
\lambda_0^{(1)} \delta_-^{-1/2} \{  \| H_1^\Lambda R_0^-
(E_0)^{1/2} \|+ \lambda_0^{(1)} \| H_2^\Lambda R_0^-
(E_0)^{1/2} \| \} \; < 1 .
\eeq
We note that $\|R_0^- (E_0)\|$ depends only on the distance from
$E_0$ to $E_-$ which is independent of $\lambda$. Consequently,
condition (\ref{lambda1}) requires that $\lambda_0^{(1)} < C_0
\delta_-$. Similarly, the operator $\{ P_+ - R_- (E_0) \mathcal{V}
(\lambda) P_+ \}$ is bounded for $| \lambda | < \lambda_0^{(1)}$.
Consequently, it follows from (\ref{feshbach1}) that the norm on the
right in (\ref{prob1}) is large if the norm of $\mathcal{G} (E_0)$
is large. To analyze this operator, note that ${\mathcal{G}}(E_0)$
can be written as
\beq\label{feshbach4}
{\mathcal{G}}(E_0)  = R_0^+ (E_0)^{ 1/2} \{ 1
+ {\tilde \Gamma}_+ (E_0) \}^{-1} R_0^+ (E_0)^{ 1/2} .
\eeq
The compact, self-adjoint operator ${\tilde \Gamma}_+ (E_0)$ has an
expansion in $\lambda$ given by
\beq\label{gamma1}
{\tilde \Gamma}_+ (E_0)  =
\sum_{j=1}^4 \lambda^j M_j ( E_0 ) ,
\eeq
where the coefficients are
given by
\bea\label{gamma2}
M_1 (E_0) & = & R_0^+ (E_0)^{ 1/2} P_+ H_{1,\omega}^\Lambda P_+
R_0^+(E_0)^{ 1/2} , \nonumber \\
M_2 (E_0) & = & R_0^+ (E_0)^{ 1/2} \{ P_+ H_{2, \omega}^\Lambda P_+
\nonumber \\
   & & - P_+ H_{1 , \omega}^\Lambda  P_- R_-(E_0) P_- H_{1 , \omega}^\Lambda
        P_+ \} R_0^+ (E_0)^{ 1/2} , \nonumber \\
M_3 (E_0) & = & -R_0^+ (E_0)^{ 1/2} \{ P_+ H_{1 , \omega}^\Lambda
P_- R_-(E_0)    P_- H_{2 , \omega}^\Lambda  P_+ \nonumber \\
  & & + P_+ H_{2 , \omega}^\Lambda P_- R_- (E_0) P_-
    H_{1,  \omega}^\Lambda P_+ \} R_0^+ (E_0)^{ 1/2} , \nonumber \\
M_4 (E_0) & = & - R_0^+ (E_0)^{ 1/2}\{ P_+ H_{2 , \omega}^\Lambda
P_- R_- (E_0) P_- H_{2 , \omega}^\Lambda P_+ \}  R_0^+ (E_0)^{ 1/2}.
       \nonumber \\
\eea

\noindent
3.The probability estimate in (\ref{prob1}) is now reduced to
\bea\label{prob2} \P \{ \mbox{dist} ( \sigma ( H_\Lambda (\lambda) ) ,
E_0 ) < \eta \}
&  = & \P \{ \; \| R_\Lambda ( E_0 ) \| > 1 / \eta \} \nonumber \\
& \leq & \P \{ \; \| {\mathcal{ G }} ( E_0 ) \| \; > 1 / ( 8 \eta ) \}
\nonumber \\
& \leq & \P \{ \; \| ( 1 + {\tilde \Gamma}_+ ( E_0 ))^{-1} \| \; >
\delta_+ (E_0) / ( 8 \eta ) \} \nonumber \\
& = & \P \{ \: \mbox{dist} ( \sigma ( {\tilde \Gamma}_+ (E_0 ) ) , -1
)
< 8 \eta / \delta_+ (E_0) \} . \nonumber \\
\eea To estimate the probability on the last line of (\ref{prob2}), we
analyze the spectrum of the operator $\tilde{\Gamma}_+ (E_0 )$. Let
$\kappa \equiv 8 \eta / \delta_+ (E_0)$ and let $E_\Lambda ( \cdot)$
be the spectral projectors for $\tilde{\Gamma}_+ (E_0 )$. Chebychev's
inequality implies that \bea\label{prob3} \P \{ \; \mbox{dist} \; (
\sigma ( \tilde{\Gamma}_+ ( E_0 ) ) , -1 ) < \kappa \} & = & \P \{ Tr
( E_{\Lambda} ( I_\kappa ) ) \geq 1 \}
\nonumber \\
& \leq & \E \{ Tr ( E_{\Lambda} ( I_\kappa ) ) \} .  \eea Let $\rho$
be a nonnegative, smooth, monotone decreasing function such that $\rho
( x ) = 1 $, for $ x < - \kappa /2 $, and $\rho ( x) = 0$, for $ x
\geq \kappa /2 $.  We can assume that $\rho$ has compact support since
$\tilde{\Gamma}_+ (E_0 )$ is lower semibounded, independent of
$\Lambda$.  As in \cite{[CHN],[HK]}, we have \bea\label{prob4}
\E_\Lambda \{ Tr ( E_{\Lambda} ( I_\kappa ) ) \} & \leq & \E_{\Lambda}
\{ Tr [ \rho ( \tilde{\Gamma}_+ (E_0) + 1 - 3 \kappa / 2 ) - \rho (
\tilde{\Gamma}_+ (E_0) + 1 + 3 \kappa / 2 ) ] \}
\nonumber \\
& \leq & \E_{\Lambda} \left\{ Tr \left[ \int_{- 3 \kappa / 2}^{ 3
      \kappa / 2} \; \frac{d}{dt} \rho ( \tilde{\Gamma}_+ (E_0) + 1 -
    t) \; dt \right] \right\} .  \eea

\noindent
4. In order to evaluate the $\rho '$ term,
we compute the action of the vector field $A_\Lambda$,
defined by
\beq\label{vf1}
A_\Lambda = \sum_{ j \in {\tilde \Lambda }} \;
\omega_j  \frac{\partial}{ \partial \omega_j } ,
\eeq
on the operator ${\tilde \Gamma}_+ (E_0)$ defined in
(\ref{gamma1})--(\ref{gamma2}).
This calculation is
carried out in \cite{[HK]}, and we obtain
%
%
%
\beq\label{vf2}
A_\Lambda {\tilde \Gamma}_+ (E_0) = {\tilde
\Gamma}_+ (E_0) + \sum_{j = 2}^6 \; \lambda^j K_j (E_0) . \eeq
The remainder terms $K_j(E_0)$ are given by
\bea\label{vf-remainder}
K_2 (E_0 ) &=&   M_2 (E_0) ,  \nonumber \\
K_3 (E_0) &=& 2 M_3 ( E_0) \nonumber \\
  & &  + R_0^+(E_0)^{1/2} \{ P_+ H_{1, \omega}^\Lambda
      R_- (E_0) H_{1, \omega}^\Lambda  R_- (E_0) H_{1, \omega}^\Lambda P_+
\}
      R_0^+(E_0)^{1/2}, \nonumber \\
K_4 (E_0) & = & 3 M_4 (E_0) + R_0^+ (E_0)^{1/2}\{ 2 P_+ H_{1,
\omega}^\Lambda
   R_- (E_0) H_{2, \omega}^\Lambda R_- (E_0) H_{1, \omega}^\Lambda P_+
  \nonumber \\
& &  +  P_+ H_{1, \omega}^\Lambda
      R_- (E_0) H_{1, \omega}^\Lambda R_- (E_0) H_{2, \omega}^\Lambda P_+
       \nonumber \\
  & & + P_+ H_{2, \omega}^\Lambda R_- (E_0) H_{1, \omega}^\Lambda R_- (E_0)
      H_{1, \omega}^\Lambda P_+  \} R_0^+ (E_0)^{1/2}, \nonumber \\
K_5 (E_0) & = &  R_0^+ (E_0)^{1/2} \{ 2 P_+ H_{1,\omega}^\Lambda R_-
(E_0)
          H_{2,\omega}^\Lambda R_- (E_0) H_{2, \omega}^\Lambda P_+ \nonumber
\\
    & & + 2 P_+ H_{2, \omega}^\Lambda R_- (E_0)H_{2, \omega}^\Lambda R_-
(E_0)
       H_{1, \omega }^\Lambda P_+  \nonumber \\
   &  & + P_+ H_{2, \omega }^\Lambda R_- (E_0) H_{1, \omega}^\Lambda R_-
(E_0)
       H_{2, \omega}^\Lambda P_+ \} R_0^+ (E_0)^{1/2} , \nonumber \\
K_6 (E_0) & = & R_0^+ (E_0)^{1/2} \{ 2 P_+ H_{2, \omega}^\Lambda R_-
(E_0)
   H_{2, \omega}^\Lambda R_- (E_0) H_{2, \omega}^\Lambda P_+ \}
   R_0^+ (E_0)^{1/2} . \nonumber \\
& &
\eea
We need to compute $\| A_\Lambda {\tilde \Gamma}_+ (E_0) \rho ' (
{\tilde \Gamma}_+ (E_0) - t + 1 ) \|$.  This requires that we choose
$|\lambda|$ sufficiently small so that
\beq\label{remainder-vf2}
\sum_{j=2}^6 \lambda^j \;
\| K_j (E_0) \| \; < \; (1 - 2 \kappa ) /2 .
\eeq
Now, by~\eqref{vf-remainder} and as $E_0\in G$ such that
$2E_0>E_++E_-$, one has
\begin{equation}
  \label{eq:11}
  \sum_{j=2}^6 \lambda^j \; \| K_j (E_0) \|\leq \frac{C\lambda^2}{\delta_+(E_0)}
\end{equation}
where $C$ depends on the gap size and on the relative $H_0$-bounds of
$H_{j, \omega}^\Lambda$ (but not on $\lambda$ is a compact interval).
Let $\lambda_0^{(2)} > 0$ be chosen so that $| \lambda | <
\lambda_0^{(2)}$ guarantees that (\ref{remainder-vf2}) holds; it
clearly suffices that $\lambda^2/\delta_+(E_0)$ be sufficiently small.
We now choose $\lambda_0 = \mbox{min} \; ( \lambda_0^{(1)} ,
\lambda_0^{(2)}
)$.\\
\noindent
5. With this choice of $\lambda_0$, we obtain the following crucial
lower bound
\beq\label{lowerbd2}
Tr \{ \rho' ( {\tilde \Gamma}_+ (E_0) + 1 - t ) A_\Lambda
{\tilde \Gamma}_+ (E_0) \}
  \geq - C_1 Tr \{ \rho' ( {\tilde \Gamma}_+ (E_0) + 1 - t ) \} ,
\eeq
for a finite constant $C_1 > 0$. Given this positivity condition,
we can finish the proof as in \cite{[CHN],[HK]}.
\end{proof}

\noindent
We need the explicit formulas in order to check the dependence of various
constants in $\lambda$ in section \ref{localization1}.
We mention that this proof implies the H{\"o}lder continuity
of the IDS outside of the spectrum of $H_0$.

\begin{cor}
\label{continuity}
Let $H_\omega ( \lambda) = H_0 + \lambda H_{1,
\omega} + \lambda^2 H_{2, \omega}$ be a random family of operators
satisfying hypotheses (H1)--(H3). Then, for any closed interval $I
\subset \R \backslash \sigma (H_0)$, there exists a constant $0 <
\lambda_0 ( I)$ such that for any $| \lambda| < \lambda_0 (I)$, the
integrated density of states for $H_\omega ( \lambda)$ on $I$ is
H{\"o}lder continuous of order $1/q$, for any $q
>1$.
\end{cor}


\section{The Proof of Localization}
\label{localization1}

We now combine the main Theorems \ref{maintheorem1} and
\ref{wegnerest1} in order to prove localization near the internal band
edges. We make the observation that for $\lambda > 0$ small, the
variation of the eigenvalues $E_j ( \lambda)$ of $H_\Lambda (\lambda)$
are $\mathcal{O} (\lambda)$. This follows by the Feynman-Hellman
Theorem since \bea\label{variation1} \frac{d E_j (\lambda)}{ d
  \lambda} &=& \langle \phi_j , \frac{d H_\Lambda (\lambda)}{d
  \lambda} \phi_j \rangle
\nonumber \\
&=& \langle \phi_j , [ H_1^\Lambda + 2 \lambda H_2^\Lambda ] \phi_j
\rangle.  \eea The second term on the last line of (\ref{variation1})
is bounded above by \beq\label{variation2} | \langle \phi_j, 2 \lambda
H_2^\Lambda \phi_j \rangle | \leq 2 \lambda \| u \|_\infty^2 .  \eeq
As for the first term, we have \beq\label{variation3} | \langle \phi_j
, H_1^\Lambda \phi_j \rangle | \leq \| H_1^\Lambda \phi_j \|, \eeq so
that using the bound $E_j < E_+$ and the fact that $H_1^\Lambda$ is
relatively $H_0^\Lambda$-bounded, we get \beq\label{variation4} \|
H_1^\Lambda \phi_j \| \leq \| H_1^\Lambda (H_0^\Lambda + 1)^{-1} \| [
E_+ + \lambda^2 \|u\|_\infty^2 + \lambda \|H_1^\Lambda \phi_j \| ],
\eeq from which it follows that for $\lambda$ small, the norm $\|
H_1^\Lambda \phi_j \| $ is uniformly bounded in $\lambda$. This result
and (\ref{variation2})--(\ref{variation4}) imply that $\frac{d E_j
  (\lambda)}{ d \lambda}$ is bounded by a constant.  It follows from
the fact that the deterministic spectrum is the union of the spectra
of the periodic approximations and from Floquet theory that the
band-edges $E_\pm ( \lambda)$ scale at most linearly in $\lambda$ as
$\lambda \rightarrow 0$. That the band edge scales at least linearly
comes from condition (Gh) (see Theorem~\ref{ids1}).

\subsection{Wegner Estimate}

We fix $0 < \lambda < \lambda_0$ and consider the Wegner estimate for
energies in the interval $I_\lambda = [ E_+ (\lambda) , E_+]$ given by
Theorem~\ref{wegnerest1}.  We note that, if we pick $E_0\in G$ such
that, $\inf(E_+-E_0,E_0-E_-)\geq\nu\lambda$ and $\lambda$ sufficiently
small (depending on $\nu$), we obtain a Wegner estimate, as in Theorem
\ref{wegnerest1}, holds true where the constant in front of
$\eta^{1/q}|\Lambda|$ in the right side of (\ref{wegner-1}) is
replaced with $C_W\lambda^{-1}$.

\subsection{Initial Length Scale Estimate}

We fix $0 < \lambda < \lambda_0$ of Theorem \ref{wegnerest1} so a
Wegner estimate holds for intervals $[ E_+ (\lambda) , E_0]$, for
any $E_0 < E_+$. We choose $L_1$, depending on $\lambda$, so that $[
E_+ (\lambda) , E_+ (\lambda) + L_1^{-1/2} ] \subset [ E_+ (\lambda)
, E_0]$. Note that $L_1 = \mathcal{O} ( \lambda^{-2})$. We now apply
Theorem \ref{initialest1}. For our fixed $\lambda$, there exist $(
\tilde{L}_0, \gamma_0)$ so that the conclusions of the theorem hold.
Consequently, taking $L_0 \equiv \sup ( \tilde{L}_0, L_1)$, we have
a Wegner estimate and an initial decay estimate (\ref{resolvent1})
for all energies in the interval $[ E_+ ( \lambda ), E_+ ( \lambda )
+ (4 L_0)^{-1/2}]$.


\subsection{Multiscale Analysis}

The multiscale analysis for the fixed energy interval $[ E_+ (\lambda) , E_+
(\lambda) + (4L_0)^{-1/2}]$
can now be performed as described, for example,
in \cite{[K1]}.


\section{Examples of Magnetic Schr{\"o}dinger Operators satisfying the
Hypotheses}
\label{example1}

In this section, we show that the hypotheses on the unperturbed
magnetic Schr{\"o}dinger operator are satisfied by many examples.
Let $\Gamma\subset\R^d$ be a nondegenerate lattice with a
fundamental cell $\mathcal{C}$. Consider $V_0: \R^d\to\R$ a smooth,
$\Gamma$-periodic potential.
We define the unperturbed periodic Schr{\"o}dinger operator $H_0$ as
\begin{equation}
  \label{eq:1}
  H_0=-\Delta+V_0 .
\end{equation}
Suppose $(E_-,E_+)$ is an open gap for $H_0$ and assume that there
is a unique Floquet eigenvalue $E( \theta)$ taking the value $E_+$
(the simplicity hypothesis (H4)). We further assume, for clarity of
presentation, that this value is taken at a single Floquet
parameter, say $\theta_0$, so that $E( \theta_0) = E_+$. Finally, we
assume that the extremum of the Floquet eigenvalue at $\theta_0$ is
quadratic nondegenerate as a function of $\theta$. This is
equivalent to hypothesis (H5), see \cite{[K2]}. Such potentials can
be constructed as sums of one dimensional operators (i.e.\ with
separate variables) or using semiclassical constructions (see
\cite{MR93c:35027} and references therein). Let $(x,\theta)\mapsto
\psi(x,\theta)$ be the smooth Floquet eigenfunction associated to
the Floquet eigenvalue $\theta\mapsto E(\theta)$ assuming the value
$E_+$ so that $E( \theta_0) = E_+$. The eigenfunction $\psi$
satisfies
\begin{equation}
  \label{eq:2}
  \begin{split}
    (H_0-E(\theta))\psi_0(\cdot,\theta)&=0\\
    \forall\gamma\in\Gamma,\ \forall x\in\R^d,\
    \psi_0(x+\gamma,\theta)&=e^{ i \gamma\theta} \psi_0(x,\theta).
  \end{split}
\end{equation}

We now consider a perturbation of $H_0$ by a $\Gamma$-periodic
vector potential $A_0 \in C^\infty ( \mathcal{C})$ with a coupling
constant $\varepsilon > 0$, and define $H_\varepsilon$ by
\beq\label{weakpert1} H_\varepsilon = ( i \nabla + \varepsilon A_0
)^2 + V_0 . \eeq For $\varepsilon$ small, we will construct the
Floquet eigenvalues and eigenfunctions of $H_\varepsilon$ near the
energy $E_+$. By the boundedness of $A_0$ and $\nabla A_0$, the
mapping $\varepsilon\mapsto H_\varepsilon$ is norm resolvent
continuous. Consequently, we know that, for $\varepsilon$ small,
there exists some $E_+(\varepsilon)$ in the spectrum of
$H_\varepsilon$ and such that $(E_-+(E_+-E_-)/2,E_+(\varepsilon))$
is in a gap of $H_\varepsilon$. Moreover, for
$|\theta-\theta_0|>\delta$, the Floquet spectrum of $H_\varepsilon$
for Floquet parameter $\theta$ is either above $E_++\delta/2$ or
below $E_-+\delta/2$; and, for $|\theta-\theta_0|\leq\delta$, there
exists a unique Floquet eigenvalue $E_\varepsilon(\theta)$ of
$H_\varepsilon$ in $[E_-+\delta/2,E_+ + \delta/2]$.
Let $c$ be a small circle around $E_+$; for
$\varepsilon$ small, the spectral projector on this Floquet eigenvalue
can be expressed as
\begin{equation}
  \label{eq:3}
  \pi_\varepsilon= \frac{1}{2 \pi i} \oint_{c}(z-H_\varepsilon)^{-1}dz,
\end{equation}
and the Floquet eigenvalue and a normalized Floquet eigenvector as
\begin{equation}
  \label{eq:4}
  \psi_\varepsilon(\theta)=\frac1{\|\pi_\varepsilon\psi_0(\theta)\|}
  \pi_\varepsilon\psi_0(\theta)\quad
  \text{and}\quad
  E_\varepsilon(\theta)=\langle
  H_\varepsilon\psi_\varepsilon(\theta),\psi_\varepsilon(\theta)\rangle.
\end{equation}
These functions are jointly real analytic in $\varepsilon$ and
$\theta$ for $\varepsilon$ near zero and $\theta$ near $\theta_0$.

We now want to consider a perturbation of $H_\epsilon$ by a vector
potential $A: \R^d\to\R^d$ supported in $\mathcal{C}$, the fundamental cell
of $\Gamma$. We recall that given $(V_0, A_0)$,
Ghribi's criterion (see~\cite{[G1]}) requires that a certain quadratic
form be sign-definite:

\begin{equation}
  \label{eq:5}
  \begin{split}
    \int_C([(A\cdot i\nabla+i\nabla\cdot A)+2\varepsilon A\cdot
    A_0]\psi_\varepsilon)(x)
    \overline{\psi_\varepsilon(x)}dx\\=2\int_C A\cdot\left[
    \text{Re}(i\nabla\psi_\varepsilon(x)\overline{\psi_\varepsilon(x)})+
    \varepsilon A_0(x) |\psi_\varepsilon(x)|^2\right]dx&\not=0.
  \end{split}
\end{equation}
So if $(V_0, A_0, \epsilon)$ are such that
\beq\label{nonvanish1}
\text{Re}(i\nabla\psi_\varepsilon(x)
\overline{\psi_\varepsilon(x)}) +\varepsilon A_0(x)
|\psi_\varepsilon(x)|^2 \not\equiv0,
\eeq
we can certainly choose vector-valued functions $A$, supported on
$\mathcal{C}$,
so that (\ref{eq:5}) is satisfied. In the applications to random
magnetic Schr{\"o}dinger operators, the function $A$ is the single-site
vector-valued function $u$ appearing in
(\ref{potential1}).

We now construct $A_0$ such that condition (\ref{nonvanish1}) is satisfied
for $\varepsilon$ small and nonzero. Notice that if
$\text{Re}(i\nabla\psi_0(x)\overline{\psi_0(x)}) \not\equiv0$, we
just need to take $\varepsilon=0$ and we can proceed with the
unperturbed periodic operator $H_0$. So we can assume that
\begin{equation}\label{vanish2}
  \text{Re}(i\nabla\psi_0(x) \overline{\psi_0(x)})\equiv0
\end{equation}
We note that if $\psi_0$ is real, then this condition
(\ref{vanish2}) holds. This is the case in the standard examples of
periodic Schr\"odinger operators having the band edge behavior that
we require. Let us analyze this condition. We prove
\begin{lemma}
  \label{le:1}
  Let $\psi_0$ satisfy $(H_0-E_+)\psi_0=0$, condition (\ref{eq:2}),
and  be such that~\eqref{vanish2}
  holds. Then, $\psi_0$ is collinear to a real-valued function. that is,
there exists a real function $f$ and a complex number $\eta$ with $|\eta| =
1$ so that $\psi_0 = \eta f$.
\end{lemma}
\noindent This in particular implies that when~\eqref{vanish2} holds,
$\theta_0$ has to be in $\frac12\Gamma^*/\Gamma^*$.\\
{\bf Proof of Lemma~\ref{le:1}.} We write $\psi_0(x)=e^{i\alpha(x)}r(x)$,
where $\alpha$ and $r$ are real valued. Clearly $\alpha$ is well
defined as soon as $r$ does not vanish. At points where $r$ is nonzero,
$r$ and $\alpha$ enjoy the same regularity as $\psi_0$.
Moreover,~\eqref{vanish2} gives
\begin{equation*}
  \nabla\alpha(x)=0\text{ if }r(x)\not=0.
\end{equation*}
Hence, $\alpha$ is constant on the connected components of the
complement of the nodal set of $\psi_0$. Fix such a component, say
$C_0$ and let $\alpha_0$ be the value taken by $\alpha$ on $C_0$.
Clearly $e^{-i \alpha_0}\psi_0 = r$ is real positive on this component. Let
$\varphi=\text{Im}(e^{- i \alpha_0}\psi_0)$. As the differential equation
in~\eqref{eq:2} has real coefficients, we see that $\varphi$ is a
solution to $(H_0-E_+)\varphi=0$. Moreover $\varphi$ vanishes on some
open set. The operator $H_0-E_+$ satisfies a unique continuation
principle. Hence, $\varphi$ vanishes everywhere which proves that
$e^{-i \alpha_0}\psi_0$ is real.\qed
\vskip.5cm
We now return to constructing $A_0$ so that (\ref{nonvanish1}) holds
assuming (\ref{vanish2}).
Differentiating in $\varepsilon$ the eigenvalue
equation $H_\varepsilon\psi_\varepsilon=E_+(\varepsilon)
\psi_\varepsilon$ (note that differentiating the boundary condition
$\psi_\varepsilon (x+\gamma,\theta)=e^{i \gamma\theta}\psi_\varepsilon
(x,\theta)$ does not change it), one computes the Taylor expansion in
$\varepsilon$ of $\psi_\varepsilon$ and gets
\begin{equation}
  \label{eq:7}
  \psi_\varepsilon=\psi_0+\varepsilon\psi'_0+O(\varepsilon^2)
\end{equation}
where
\begin{equation}
  \label{eq:8}
    \psi'_0=-(H_0-E_+)^{-1}(1-\pi_0)(A_0\cdot i\nabla+i\nabla\cdot
A_0
)\psi_0.
\end{equation}
The Taylor expansion~\eqref{eq:7} can be differentiated in $x$ and
$\theta$ (for $\theta$ close to $\theta_0$).\\
Substituting~\eqref{eq:7} into~\eqref{eq:5}, and taking~\eqref{vanish2} into
account, we see that we need to find $A_0$ such that
\begin{equation}
  \label{eq:9}
  \text{Re}(i\nabla\psi'_0(x)
  \overline{\psi_0(x)})+\text{Re}(i\nabla\psi_0(x)
  \overline{\psi'_0(x)})+A_0(x)|\psi_0(x)|^2\not\equiv0.
\end{equation}

We will now construct $A_0$ not identically vanishing such that
$\psi'_0$ vanishes identically. Hence,~\eqref{eq:9} is satisfied.
Consider $\nabla^\perp$ a constant coefficient vector field such that
\begin{equation*}
  \nabla\cdot\nabla^\perp=0
\end{equation*}
where, as above, $\cdot$ denotes the scalar product in $\R^d$.
For example, in dimension $d=2$, we choose $\nabla^\perp=(-\partial_2,
\partial_1)$. In general, the $i^{th}$ component of $\nabla^\perp$
has the form $M_{ij} \partial_j$ for a constant real skew symmetric matrix
$M_{ij} = - M_{ji}$.
Notice that for any differentiable function $\psi:\
\R^d\to\C$, one also has
\begin{equation}
  \label{eq:10}
  \nabla(\psi)\cdot\nabla^\perp(\psi)= -
\nabla^\perp(\psi)\cdot\nabla(\psi)=
  \frac12\nabla\cdot\nabla^\perp(\psi^2)-\psi\nabla\cdot\nabla^\perp(\psi)=0.
\end{equation}
As $\psi_0$ is not a constant, we can choose $\nabla^\perp$ so that
$\nabla^\perp\psi_0$ does not vanish identically. Then, we set
$A_0=\nabla^\perp(\psi_0^2)$. As $V_0$ is infinitely differentiable,
the eigenvector $\psi_0$ is too, and this implies that $A_0$ is
infinitely differentiable.  Moreover, as $\theta_0\in
\frac12\Gamma^*/\Gamma^*$, one has
\begin{equation*}
  A_0 (x+\gamma)=\nabla^\perp(\psi_0^2(x+\gamma))
  =e^{2 i \gamma\theta_0}\nabla^\perp(\psi_0^2(x))=\nabla^\perp(\psi_0^2(x))
  =A_0(x).
\end{equation*}
Finally, using~\eqref{eq:10}, we compute
\begin{equation*}
    (A_0 \cdot i\nabla+i\nabla\cdot A_0)\psi_0=
4i\nabla(\psi_0)\cdot\nabla^\perp(\psi_0)\psi_0
    +2i\nabla\cdot\nabla^\perp(\psi_0)\psi^2_0=0.
\end{equation*}
Hence, $\psi'_0=0$. This completes the construction of the example.
To summarize: Given a smooth periodic electrostatic potential $V_0$,
we can compute a smooth periodic vector potential $A_0$ and a
single-site vector potential $u=A$ satisfying (H2) and (\ref{eq:5})
so that the random Schr{\"o}dinger operator $H_\omega ( \lambda,
\epsilon) = ( -i \nabla - \epsilon A_0 - \lambda A_\omega)^2 + V_0$,
for $\epsilon> 0$ and small, satisfies the condition of Ghribi
(\ref{eq:5}). Consequently, the IDS exhibits Lifshitz tails behavior
at the inner band-edges.
\par Let us notice here that, by Lemma~\ref{le:1}, if we hope to
construct an example of a periodic Schr{\"o}dinger operator without a
periodic magnetic field satisfying (\ref{eq:5}), it is necessary and
sufficient to find a periodic Schr{\"o}dinger operator that the energy
$E(\theta)$ reaches a band-edge for a Floquet parameter not belonging
to $\frac12\Gamma^*/\Gamma^*$ and for which the band edge is simple.
To our knowledge, no such example is known. In dimension $d=1$, it is
known that this can not happen. In larger dimensions, the difficulty
in producing such an example can be understood as a consequence of the
fact that the points in $\frac12\Gamma^*/\Gamma^*$ are always critical
points for simple Floquet eigenvalue. This clearly makes the
construction of such an example by perturbation theory difficult.


\small \noindent {\sc D\'epartement de Math\'ematiques \\
Site Saint-Martin 2 BP 222\\
 Université de Cergy-Pontoise \\
 95302 Cergy Pontoise Cedex \\
 France \\
 e-mail:{\tt fatma.ghribi@u-cergy.fr} }

\vspace{2mm}

\noindent
{\sc Department of Mathematics\\
University of Kentucky\\
Lexington, Kentucky 40506--0027\\
USA\\
e-mail: {\tt hislop@ms.uky.edu}}

\vspace{2mm}

\noindent
{\sc L.\ A.\ G.\ A. \\
Institut Galil{\'e}e \\
Universit{\"e} Paris-Nord \\
F-93430 Villetaneuse \\
France \\
et\\
Institut Universitaire de France \\
e-mail:{\tt klopp@math.univ-paris13.fr} }

\end{document}